\documentclass[letterpaper]{article}
\usepackage[english]{babel}

\usepackage{amsmath}
\usepackage{graphicx}
\usepackage[colorlinks=true, allcolors=blue]{hyperref}

\usepackage{algorithm}
\usepackage{subcaption}
\usepackage{csquotes}
\usepackage{amsthm}
\theoremstyle{definition}
\newtheorem{definition}{Definition}

\usepackage{algorithmicx}
\usepackage{algpseudocode}
\algdef{SE}[DOWHILE]{Do}{doWhile}{\algorithmicdo}[1]{\algorithmicwhile\ #1}%
\usepackage{amsmath, amssymb}
\usepackage{amsthm}
\usepackage{graphicx}
\usepackage{pgfplots}
\usepackage{tikz}
\usepackage{authblk}
\usetikzlibrary{positioning}
\usetikzlibrary{fit,positioning,arrows}
\tikzstyle{arrow} = [thick,->,>=stealth]
\tikzstyle{arrow2} = [dashed,->,>=stealth]
\tikzstyle{arrow3} = [thick,->,>=stealth, red]
\tikzstyle{s1} = [circle, fill=white!90,rounded corners, minimum width=1cm, minimum height=1cm,text centered, draw=black]
\usepackage{pdflscape}
\usetikzlibrary{matrix}
\usepackage{import}
\DeclareMathOperator*{\argmax}{arg\,max}

\usetikzlibrary{decorations.markings, arrows.meta}

\tikzset{
    marrow/.style={decoration={markings,mark=at position 0.5 with {\arrow{#1}}}, postaction=decorate, thick}
}

\title{Fractal Conditional Correlation Dimension Infers Complex Causal Networks}

\author[1, 2, 3, *]{Özge Canlı Usta}
\author[1, 2, *]{Erik M. Bollt}

\affil[1]{Department of Electrical and Computer Engineering, Clarkson University, 8 Clarkson Ave., Potsdam, NY 13699, USA.  }
\affil[2]{ Clarkson Center for Complex Systems Science, Clarkson University, 8 Clarkson Ave., Potsdam, NY 13699, USA.}
\affil[3]{Department of Electrical and Electronics Engineering,  Dokuz Eylül University, Izmir, 35390, Turkey }
\affil[*]{\texttt{ocanlius@clarkson.edu} ~ \texttt{bolltem@clarkson.edu} }

\date{}

\begin{document}
\maketitle

\begin{abstract}
\noindent Determining causal inference has become popular in physical and engineering applications. While the problem has immense challenges, it  provides a way to model the complex networks by observing the time series. In this paper, we present the optimal conditional correlation dimensional geometric information flow principle ($oGeoC$) that can reveal direct and indirect causal relations in a network through geometric interpretations. 
We introduce two algorithms that utilize the $oGeoC$ principle to discover the direct links and then remove indirect links. The algorithms are evaluated using coupled logistic networks. The results indicate that when the number of observations is sufficient, the proposed algorithms are highly accurate in identifying direct causal links and have a low false positive rate.
\end{abstract}

\section{Introduction}

Causal inference has attracted attention in various scientific fields, from engineering \cite{sudu2016information} to climate science \cite{runge2019inferring, runge2023causal} and from neuroscience \cite{seth2013granger} to ecological systems \cite{sugihara2012detecting}. The problem is reconstructing the causal relations from the observed time series of a complex network. However, the underlying dynamics of the networks are often unknown, and the observations can be limited. Hence, the ability to model the networks and infer causal relationships among the systems can be quite challenging.

We have written that \cite{bollt2018open, surasinghe2020geometry}, \enquote{a basic question when defining the concept of information flow is to contrast versions of reality for a dynamical system. Either a subcomponent is closed or alternatively, there is an outside influence due to another component}.
Claude Granger’s Nobel prize \cite{hendry2004nobel}-winning work leading to Granger Causality (see also Wiener \cite{wiener1956theory}) formulates causal inference as a concept of quality of forecasts. That is, we ask, does system $X$ provide sufficient information regarding forecasts of future states of system $X$, or are there improved forecasts with observations from system $Y$? We declare that $X$ is not closed, as it is receiving influence (or information) from system Y, when data from $Y$ improve forecasts of $X$, and this is called Weiner-Granger causality, WGC.
In Granger's original test for causality (GC) between two time series, a time series $Y$ has a causal inference on a second time series $X$ if the future of $X$  includes information from past terms of $Y$ \cite{granger1969investigating} as decided by forecasting $X$ in two different ways with linear models, with and without considering the information from $Y$.
GC deals with the identification of causality in stochastic and linear systems, and its extensions have been introduced to tackle the problem of detecting causation in separability between multivariate-time series and nonlinear models \cite{marinazzo2008kernel, barrett2010multivariate, marinazzo2011nonlinear}. 
Other variations on the concepts of WGC exist based on other concepts of forecasting.

Cross-mapping (CM) techniques, which use the predictions of one system based only on the past observations from the other system, are also employed in detecting causal inference problems \cite{rulkov1995generalized, schiff1996detecting}. 
Rulkov et al. have studied the connections of two unidirectional coupled systems and the detection of synchronization using CM-based technique in \cite{rulkov1995generalized}. The authors have also focused on the connections of unidirectional coupled systems and applied the mutual nonlinear prediction method to neuroscience \cite{schiff1996detecting}. Several methods have been proposed
to infer causal relationships and synchronization using CM techniques \cite{ arnhold1999robust, quiroga2002performance, andrzejak2003bivariate, chicharro2009reliable}. 
Following the line of CM techniques-based research works, Sugihara et al. have also proposed convergent cross-mapping (CCM)  
that utilizes a state-space reconstruction technique \cite{sugihara2012detecting}. CCM can identify causality in weakly coupled networks and find causal links in complex ecosystems. Although CCM requires a large amount of data and fails in case of strong coupling or synchrony, CCM and its alternatives have also been widely used in recent years \cite{ye2015distinguishing, monster2017causal, breston2021convergent}. 

On the other hand, information-theoretic approaches are implemented to solve the causal inference problem in many applications due to being model-free, including transfer entropy \cite{schreiber2000measuring, sun2014causation, sun2015causal, lord2016inference}. Of particular interest to us here is the more nuanced concept of  direct information flow, which considers if $X$ causes $Y$ conditioned on considering intermediaries $Z$:  that is, if $X$ flows through $Z$ to influence $Y$, but perhaps there is no direct influence from $X$ to $Y$.
In a particular study, Sun et al. have suggested the optimal causation entropy principle (oCSE), an algorithm that reveals the causal relations in a complex network by using causation entropy \cite{sun2015causal},  to learn direct and indirect influences. The principle is based on the idea that the causal parents of a node in the network contain the minimal set of nodes that maximize causation entropy. The oCSE principle allows us to differentiate causal parents of a node from indirect influences of a node by using discovery and removal algorithms. 

The idea of understanding connections between geometric information flow and causal inference is investigated in recent decades \cite{janjarasjitt2008approach, krakovska2013interdependence, krakovska2015causality, surasinghe2020geometry, krakovska2019correlation}.
An index has been proposed in \cite{janjarasjitt2008approach} where it was called the dynamic complexity coherence measure. The index is the ratio of the sum of the correlation dimensions of individual subsystems to the correlation dimension of the coupled dynamical system. If the two systems are independent, the sum of the correlation dimensions of individual subsystems equals the correlation dimension of the concatenated dynamical system. However, if two systems are coupled, the sum of the correlation dimensions of individual subsystems is greater than the correlation dimension of the concatenated dynamical system. The index can determine the degree of synchronization and the presence of coupling \cite{janjarasjitt2008approach}. 
The authors in \cite{krakovska2013interdependence} have also shown that the correlation dimension can reveal the presence and the direction of coupling. The synchronization of coupled systems can be determined by using this method.

Krakovsk\'a has used the correlation dimension to detect causality \cite{krakovska2019correlation}. The study investigates the causal relevance between or within two systems for the different coupling strengths using the correlation dimension in the reconstructed state space. 
It has been emphasized that the correlation dimension in causal analysis can be a promising method between and within systems. Furthermore, the study states that correlation-based methods provide some advantages in finding causal relations in dynamical systems when we have sufficiently long observations and the states are observable.

Surasinghe and Bollt have suggested the correlation dimension geometric information flow measure to quantify causal inference between two related systems in the geometric sense \cite{surasinghe2020geometry}. The authors have proposed a new measure, i.e., geometric information flow $GeoC_{\cdot \rightarrow \cdot}$, based on conditional correlation dimension, which enables the identification of causal relations between two related systems by geometric terms. They found that $GeoC_{\cdot \rightarrow \cdot}$ provides us with geometric interpretable results when detecting causality in synthetic and real examples.

Conversely, Cummins et al. have established a theoretical model that builds on Takens’ theorem \cite{takens1981detecting} for recovering dynamic interactions between weekly coupled or moderately coupled in dynamical systems. The authors have examined the limitations of the state-space reconstruction methods. The manifold of the systems from one single coordinate observation function has been reconstructed. Then, the approach seeks to identify the reconstructions that have mutual driving, one-directional driving, or are completely independent.
The approach fails to recover self-loops, and cannot differentiate between mutual and unidirectional dynamical driving in connected components \cite{cummins2015efficacy}.
Although the study in \cite{krakovska2019correlation} claims that the correlation dimension reveals the causal relations in the reconstructed space, it can fail in some cases. If two uncoupled systems $(X, Y)$ are driven by a hidden common driver $Z$, $X$ and $Z$ cannot distinguish, and it implies a directional link from $X$ to $Y$ when $X$ and $Z$ are synchronized. 

Although the correlation dimension and measures based on it have been explored for detecting causality and synchronization, they have not been studied extensively to reveal connections in the network. The existing methods in the literature are particularly interested in detecting synchronization.
In this paper, we focus on detecting causality in the networks, and assuming the networks are not synchronized.

Additionally, the previously discussed CM or state-space methods reconstruct the phase space from a single observation of a node. In contrast, we observe all states of the subsystems in the network. Even if we analyze the correlation dimension in the reconstructed state-space or observe all states, using only the correlation dimension may be insufficient to determine the direct and indirect influences of the network. However, we have utilized the conditional correlation dimensions-based measure in \cite{surasinghe2020geometry} to infer direct and indirect influences in this paper.

The main goal of this paper is to quantify the causal inference between the subsystems in a network in the geometric sense.  Unlike previous studies, this paper extends the analysis of causal inference problems using only geometric interpretation to detect causal links in the networks. Expanding upon the fractal geometric concepts of the consequence of information flow in \cite{surasinghe2020geometry}, we develop optimal conditional correlation dimensional geometric information flow principle ($GeoC$) that resembles the oCSE principle previously proposed by Sun et al. \cite{sun2015causal}. We present two algorithms to detect the causal links and remove indirect links using the correlation dimension geometric information flow $GeoC$.  
The performance of the $oGeoC$ algorithm is investigated in coupled logistic networks.

\section{Problem Statement}

\subsection{Preliminaries and Basic Definitions}
 In this section, we present the notation and the basic definitions. A graph, $\mathcal{G} \equiv (\mathcal{V}, \mathcal{E})$ is defined by the set of vertices (nodes), $\mathcal{V} = \left\{ v_1 , v_2, \ldots, v_N \right\}$, and the set of edges (links), $\mathcal{E} \subseteq \mathcal{V} \times \mathcal{V}$. 
If $\forall$ $(v_i, v_j) \in \mathcal{E} \implies (v_j, v_i) \in \mathcal{E}$, the graph is undirected. Otherwise, it is defined as a directed graph. The set of $\mathcal{N}_i = \{ v_j \in \mathcal{V} | (v_i, v_j) \in \mathcal{E} \}$ is denoted as the parents of $i^{th}$ node. In short, we denote $v_i$ simply as $i$. The graphs can also be represented by their adjacency matrix $\mathbf{A}$. The elements of $\mathbf{A}$ are $a_{ij} =1$, if there is an edge from $j$ to $i$. Otherwise, $a_{ij} = 0$.

Consider a discrete-time dynamical system in $\mathbb{R}^d$ expressed as 
\begin{equation}
	\mathbf{x}_{n+1} = f(\mathbf{x}_n)
\end{equation}
where $	\mathbf{x}_n  \in   \mathbb{R}^d$ is the state variable  at time step $n$ and $f(\cdot):\mathbb{R}^d \rightarrow \mathbb{R}^d $ is the local dynamics.  We also consider a discrete-time dynamical network consisting of $N$ identical components

\begin{equation}
	\mathbf{x}_{n+1}^{(i)} = f(\mathbf{x}_{n}^{(i)}) + \sigma \sum_{\substack{i=1 \\ i\neq j}}^N a_{ij} \boldsymbol{\kappa} g(\mathbf{x}_{n}^{(i)}, \mathbf{x}_{n}^{(j)})    \quad i = 1, 2, \ldots, N. 
 \label{eq:network_equation}
\end{equation}
Here, $	\mathbf{x}_{n}^{(i)} \in   \mathbb{R}^d $ is the state variable of node $i$ at time step $n$, $f(\cdot):\mathbb{R}^d \rightarrow \mathbb{R}^d $ is the local dynamic, $\sigma $ is the coupling strength, $a_{ij}$ represents the coupling from  node $j$ to node $i$ and it can be expressed in matrix form $\mathbf{A} \in \mathbb{R}^{N \times N} $, $\boldsymbol{\kappa} \in \mathbb{R}^{d \times d}$ is the inner coupling matrix and $g(\mathbf{x}_{n}^{(i)}, \mathbf{x}_{n}^{(j)}): \mathbb{R}^d \times \mathbb{R}^d  \rightarrow \mathbb{R}^d $ is the coupling function.  To simplify the notation, we define the next step of $\mathbf{x}_{n}^{(i)}$ as, 
\begin{equation}
\mathbf{x}^{'(i)} = \mathbf{x}_{n+1}^{(i)},
\end{equation}
where the $'$ denotes the next time step, as an alternative notation to explicitly indexing time.

Let $\left\{\mathbf{x}_{n}^{(i)} \right\}_{n=1}^T $ and $\left\{\mathbf{x}_{n}^{(j)} \right\}_{n=1}^T $ represent sets of measurements from a network in \eqref{eq:network_equation}. Assume that a manifold of observations of $(\mathbf{x}_{n}^{(i)}, \mathbf{x}_{n}^{(j)}, \mathbf{x}_{n}^{'(i)})  \in (\mathbf{X}^{(i)} \times \mathbf{X}^{(j)}  \times \mathbf{X}^{'(i)} $) and $(\mathbf{x}_{n}^{(i)}, \mathbf{x}_{n}^{'(i)})  \in (\mathbf{X}^{(i)}  \times \mathbf{X}^{'(i)} $) are defined as $\mathcal{M}$ and $\Omega_1$, respectively. Based on how these manifolds lie in the space provides crucial information about whether $\mathbf{x}_{n}^{'(i)}$ depends only on $\mathbf{x}_{n}^{(i)} $ or on $(\mathbf{x}_{n}^{(i)}, \mathbf{x}_{n}^{(j)})$. Thus, using the dimensions of the manifold of the subsystems can be decisive in determining causal inference between systems \cite{surasinghe2020geometry}. First, the conditional correlation dimensional geometric information flow is defined as follows: 

\begin{definition}{(Conditional Correlation Dimensional Geometric Information Flow \cite{surasinghe2020geometry}).} Assume that $\mathcal{M}$ and  $\Omega_1$ are bounded Borel sets. Let $\mathcal{M}$ be a manifold of data set taken at time steps from $1$ to $T+1$ for node $i$  as $(X_1^{(i)}, X_2^{(i)}, \ldots, X_T^{(i)}, X_{T+1}^{(i)})$ and let $\Omega_1$ be a set $\mathbf{X^{(i)}}=(X_1^{(i)}, X_2^{(i)}, \ldots, X_T^{(i)})$ taken at time steps from $1$ to $T$ for node $i$. The geometric information flow $Geo(\cdot | \cdot)$ is defined in the sequel: 
	\begin{equation}
		Geo(X^{'(i)} | \mathbf{X^{(i)}})  = \mathcal{D}_2(\mathcal{M}) - \mathcal{D}_2(\Omega_1)
		\label{eq:conditional_correlation_dimensional_geo}
	\end{equation}
\end{definition}
\noindent where $\mathcal{D}_2(\cdot)$ is the correlation dimension of the given dataset \cite{kantz2003nonlinear}. Then,  the authors in \cite{surasinghe2020geometry} have defined 
correlation dimensional geometric information flow between two systems by the following:

\begin{definition}{(Correlation Dimensional Geometric Information Flow \cite{surasinghe2020geometry})} Consider $\mathbf{X}^{(i)}=(X_1^{(i)}, X_2^{(i)}, $ $\ldots, X_T^{(i)})$ and $\mathbf{X}^{(j)}=(X_1^{(j)}, X_2^{(j)}, \ldots, X_T^{(j)})$ as time series measured at time steps from $1$ to $T$ for nodes $i$ and $j$, respectively. The correlation dimensional geometric information flow from $j$ to $i$ is measured by using conditional correlation dimension in \eqref{eq:conditional_correlation_dimensional_geo} and given by 
	\begin{equation}
		GeoC_{j \rightarrow i} := Geo(X^{'(i)} | \mathbf{X}^{(i)} ) - Geo(X^{'(i)}  | \mathbf{X}^{(i)}, \mathbf{X}^{(j)}).
  \label{eq:geoC_j_i}
	\end{equation}
\end{definition}

It is clear that, if $j$ influences $i$, then $	GeoC_{j \rightarrow i}  > 0$. However, if $j$ does not influence $i$, $	GeoC_{j \rightarrow i}  = 0$. 

$GeoC$ is based on quantifying information flow between variables and how manifolds are mapped \cite{surasinghe2020geometry}.
The study also investigates information flow between two systems, even if the observation set is not only a manifold but also a fractal. It lies on how the fractal dimension changes through the transformations \cite{sauer1991embedology}. 

\subsection{Geometric Causation of Information Flow in Networks}

We aim to extend the previous concept of correlation dimension geometric information flow to the networks. The idea is to make an analogy between the oCSE principle \cite{sun2015causal} and $GeoC$ \cite{surasinghe2020geometry}. Hence, the extension of the proposed geometric measure $GeoC_{j \rightarrow i}$ leads us to solve this problem.

\begin{definition}{(optimal Conditional Correlation Dimensional Geometric Information Flow ($oGeoC$))} Let  $I$, $J$, and $K$ be subsets of nodes in a network. The correlation dimensional geometric information flow from $J$ to $I$ by conditioning on $K$ is defined as 
\begin{equation}
    GeoC_{J \rightarrow I | K} := Geo(X^{'(I)} | \mathbf{X}^{(K)} ) - Geo(X^{'(I)}  | \mathbf{X}^{(J)}, \mathbf{X}^{(K)})
    \label{eq:geo_j_i_k}
\end{equation}
\label{def:geoc_network}
\end{definition}
\noindent where $X^{'(I)}$ is the observation at $T+1$ for subset $I$ and $\mathbf{X}^{(J)}=(X_1^{(J)}, X_2^{(J)}, \ldots, X_T^{(J)})$ and $\mathbf{X}^{(K)}=(X_1^{(K)}, X_2^{(K)}, \ldots, X_T^{(K)})$  are time series measured at time steps from $1$ to $T$ for nodes $J$ and $K$, respectively. It is clear that if $\mathbf{X}^{(I)} = \mathbf{X}^{(K)} $, then \eqref{eq:geo_j_i_k} becomes, 

\begin{equation}
     GeoC_{J \rightarrow I | K}  =   GeoC_{J \rightarrow I | I} =  GeoC_{J \rightarrow I }
\end{equation}
as in denoted \eqref{eq:geoC_j_i}. Moreover, if $\mathbf{X}^{(K)} = \emptyset$, $GeoC_{\cdot \rightarrow \cdot ~|~  \cdot}$ simplifies to, 

\begin{equation}
   GeoC_{J \rightarrow I | \emptyset}  = Geo(X^{'(I)}) - Geo(X^{'(I)}  | \mathbf{X}^{(J)}).
   \label{eq:geoc_emptyset}
\end{equation}

\noindent Consider the case where $ J \subset K$. In this case, $Geo(X^{'(I)}  | \mathbf{X}^{(J)}, \mathbf{X}^{(K)})$ reduces to $Geo(X^{'(I)}  | \mathbf{X}^{(J)})$, which implies that $GeoC_{J \rightarrow I | K} = 0$. Furthermore, if $J \subset \mathcal{N}_I$ and $J \not \subset K $, then $GeoC_{J \rightarrow I | K}  > 0 $.

Using Definition \ref{def:geoc_network} and its properties, 
we can  quantify information flow between variables geometrically. Moreover, we can identify direct influences and indirect influences by using similar algorithms that were earlier designed in \cite{sun2015causal}. \enquote{FORWARD GEOC} in Algorithm \ref{alg:forword_backward_geoc_algorithm} computes $GeoC_{j~ \rightarrow ~ i ~|~ \mathcal{K}} $ for each node $i \in \mathcal{V}$, finds the maximum of $GeoC_{j~ \rightarrow ~ i ~|~ \mathcal{K}} $ over $j$. Then, the algorithm discovers one of the causal parents of $i$ in each iteration and updates $\mathcal{K}$ set iteratively that until $GeoC_{j~ \rightarrow ~ i ~|~ \mathcal{K}} $ reaches to zero $(\varepsilon_{forward})$. In \enquote{BACKWARD GEOC}, the candidate set of the causal parents of $i$ is used and the algorithm calculates $GeoC_{j~ \rightarrow ~ i ~|~ \mathcal{K} -\{j \}} $ over $K$ set. If each $GeoC$ is zero ${(\varepsilon_{backward})}$, then the candidate set of causal parent of $i$ is removed from $K$ set. \enquote{BACKWARD GEOC} also returns the estimated adjacency matrix $\mathbf{\hat{A}}$ to present the graph.  

$\varepsilon_{forward}$ and $\varepsilon_{backward}$ can be determined by setting a threshold or performing a significant test statistically \cite{good2015}. 
We utilized a shuffle test to determine the zero $(\varepsilon_{forward})$ in Algorithm \ref{alg:forword_backward_geoc_algorithm}.  The shuffle test procedure is presented in Appendix \ref{sec:shuffle_test_to_determine_the_zero}.  We set a threshold for $\varepsilon_{backward}$ in the backward algorithm. 
Note that the computation of the correlation dimension is required in Algorithm \ref{alg:forword_backward_geoc_algorithm}. Hence, the estimation of the correlation dimension is explained in Section \ref{sec:estimation_of_correlation_dimension}.


\begin{algorithm}[!ht]
\caption{$oGeoC$ algorithm}
\begin{algorithmic}[1]
\Procedure{Forward GeoC}{$\mathbf{x}_n^{(i)}~ \forall~ i= 1, 2, \ldots N$ and $n=1, 2, \ldots, T$ ; \newline
$\mathcal{V}$: the vertex set; $\varepsilon_{forward}$: threshold for zero}
\State \textit{Initialize:} $\mathcal{N} \gets \emptyset $
 \For {$i \in \mathcal{V}$} 
 \State \textit{Initialize:} $\mathcal{K} \gets \emptyset $, $index\_max \gets \emptyset $
 \Do \{$\mathcal{K}  \leftarrow \mathcal{K} \cup ~ \{ index\_max  \} $ \}
 \For {$j \in \mathcal{V} -\mathcal{K}$}
\State geoc\_j[j] =  $GeoC_{j~ \rightarrow ~ i ~|~ \mathcal{K}} $ 
 \EndFor
\State $max_{geoc} = \max_{j} \{geoc\_j \}$  
 \State  $index\_max =  \argmax \{geoc\_j \}$
    \doWhile{ $max_{geoc} > \varepsilon_{forward}$} \label{end_of_do} 
    \State $\mathcal{N} [i] \leftarrow \mathcal{K}$ 
  \EndFor

    \State \textbf{return} $\mathcal{I}$, $\mathcal{N}$ 
\EndProcedure

\Procedure{Backward GeoC}{$\mathbf{x}_n^{(i)}~ \forall~ i= 1, 2, \ldots N$ and $n=1, 2, \ldots, T$ ; \newline
Set of nodes $\mathcal{I} \subset \mathcal{V}$; and set of nodes $\mathcal{N} \subset \mathcal{V}$ :the candidate set of causal parents of $\mathcal{I}$; $\varepsilon_{backward}$: threshold for zero}
        \\ \textit{Initialize:} $\mathbf{\hat{A}} = \mathbf{0}_{N \times N }$ whose elements $[\hat{a}_{ij}]$
   \For {$i \in \mathcal{I}$}
   \State $\mathcal{K} \gets \mathcal{N}[i]$
\For {$j \in \mathcal{K}$}
  \If {$GeoC_{j~ \rightarrow ~ i ~|~ \mathcal{K} -\{j \}}  >  \varepsilon_{backward} $}
  \State $\hat{a}_{ij} = 1$
  \EndIf
 \EndFor
   \EndFor
    \State \textbf{return} $\mathbf{\hat{A}}$: estimated adjacency matrix
\EndProcedure
\end{algorithmic}
\label{alg:forword_backward_geoc_algorithm}
\end{algorithm}

\subsubsection{Estimation of Correlation Dimension}
\label{sec:estimation_of_correlation_dimension}
$\mathcal{D}_2(\cdot)$ is used to estimate $GeoC_{J \rightarrow I | K}$. Hence, the correlation dimension and its implementation details are discussed in this section. 

Consider the probability of a trajectory being within a ball $\mathcal{B}_\epsilon(\mathbf{x})$ of radius $\epsilon$ around $\mathbf{x}$, defined as $p_{\epsilon}(\mathbf{x}) = \int_{\mathcal{P}_{\epsilon{(\mathbf{x}})}} d \mu (\mathbf{x}) $. Then, the generalized correlation integral becomes \cite{kantz2003nonlinear}
\begin{equation}
C_q(\epsilon) =  \int_{\mathbf{x}} p_{\epsilon}(\mathbf{x})^{q-1} d \mu (\mathbf{x}).
\label{eq:generalized_integral}
   \end{equation}
The correlation integral in Equation \eqref{eq:generalized_integral} can be rewritten using a Heaviside step function as follows 

\begin{equation}
    C_q(\epsilon) = \int_{\mathbf{x}} \left[ \int_{\mathbf{y}} \Theta(\epsilon - ||\mathbf{x} -\mathbf{y} ||) \,d \mu(\mathbf{y}) \right]^{q-1} \,d \mu(\mathbf{x}).
    \end{equation}
Here, $\Theta$ is Heaviside step function, defined as $\Theta(x) = 1$ if $x > 0 $ and $\Theta(x) = 0$ otherwise. 

Grassberger and Proccacia have discussed the correlation integral for the case of $q=2$ \cite{grassberger1983characterization, grassberger1983measuring}. From a finite set of observations of $\mathbf{x}_i$, the estimation of the correlation integral is given as

\begin{equation}
     \hat{C}(\epsilon) =  \dfrac{1}{[T(
     T-1)]^{q-1}} \sum_{i=1}^T \left[ \sum_{i \neq j} \Theta(\epsilon - ||\mathbf{x}_i -\mathbf{x}_j ||) \right]^{q-1}
\end{equation}

where $T$ is the number of samples. The modified version of the correlation sum when $q=2$ is described in \cite{theiler1987efficient}

\begin{equation}
     \hat{C}(\epsilon) =  \dfrac{2}{T(T-1) } \sum_{i=1}^T  \sum_{j \neq i+1}^T \Theta(\epsilon - ||\mathbf{x}_i -\mathbf{x}_j ||).
     \label{eq:correlation_sum}
\end{equation}

The summation terms count the pairs
 $(\mathbf{x}_i, \mathbf{x}_j)$ for which distance $||\mathbf{x}_i -\mathbf{x}_j || < \epsilon $. It is expected that $ \hat{C}(\epsilon)$ scales a power law, $\hat{C} \propto \epsilon^{\mathcal{D}_2} $ when $N \rightarrow \infty$ and $\epsilon \rightarrow 0$.  The correlation dimension $\mathcal{D}_2$ is defined as
 \begin{equation}
 \begin{aligned}
     d(T, \epsilon) = \dfrac{\partial \ln C(\epsilon, T)}{\partial \ln \epsilon} \\
 \mathcal{D}_2 = \lim_{\epsilon \rightarrow 0} \lim_{N \rightarrow \infty} d(T, \epsilon).
  \end{aligned}
 \end{equation}

A well-known technique for estimating $\mathcal{D}_2$ involves obtaining the slope of $\ln C(\epsilon, T)/ \ln \epsilon $ curve in linear regions for $T \gg 0 $ \cite{abarbanel2012analysis}. First, 
$\ln C(\epsilon, T)$ is plotted against $\ln \epsilon$  by increasing $\epsilon$ until $\ln C(\epsilon, T)$ no longer changes with increasing $\ln \epsilon $. Then, the slope of $\ln C(\epsilon, T)/ \ln \epsilon $  in the linear region is determined using a numerical estimation method, particularly least squares estimation. Clearly,  the estimation of $\mathcal{D}_2$ depends on the number of samples $T$, minimum radius $\epsilon_{min}$,  maximum radius $\epsilon_{max}$ and the number of radius steps $\#_{rs}$ between $\epsilon_{min}$ and $\epsilon_{max}$. The details are demonstrated in Appendix \ref{sec:correlation_dim}.

\section{Results}

In this section, we present some examples to demonstrate the performance of the proposed method. 

\begin{figure*}[!ht]
    \centering

        \centering
            \centering
            \begin{subfigure}[b]{0.41\textwidth}
                \centering
\includegraphics[scale=0.8]{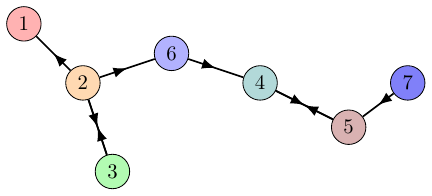}
                \caption{ }
                \label{fig:directed_7_nodes}
            \end{subfigure}
            \hfill
            \begin{subfigure}[b]{0.41\textwidth}
                \centering
\includegraphics[scale=0.8]{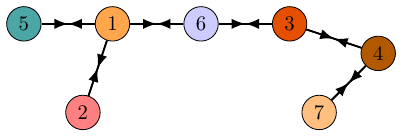}
                \caption{}
                \label{fig:bidirected_7_nodes}
            \end{subfigure}
       
            \centering
            \begin{subfigure}[b]{0.41\textwidth}
                \centering   \includegraphics[scale=0.8]{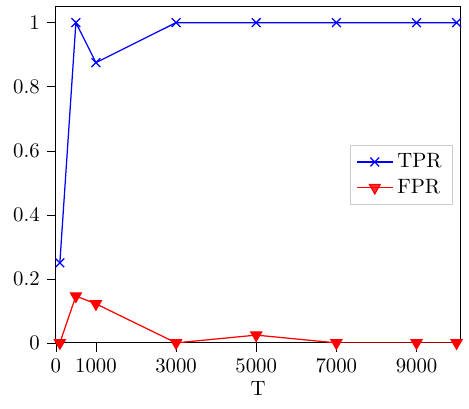}
                \caption{}             \label{fig:directed_tpr_fpr_7_nodes}
            \end{subfigure}
            \hfill
            \begin{subfigure}[b]{0.41\textwidth}
                \centering  
  \includegraphics[scale=0.8]{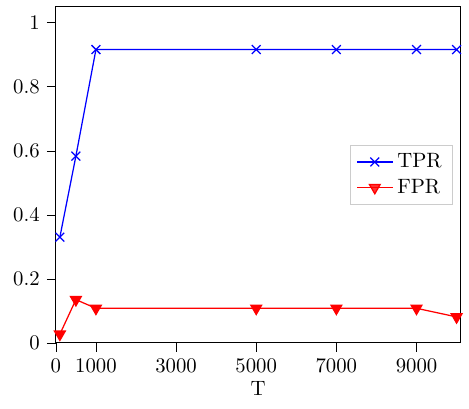}
                \caption{}             \label{fig:bidirected_tpr_fpr_7_nodes}
            \end{subfigure}

    \caption{The performance of the proposed algorithm for the networks in Figures \ref{fig:directed_7_nodes} and \ref{fig:bidirected_7_nodes}.  (a) A network  with directed coupling consisting of $N=7$ nodes and 8 links. (b) A network with bidirectional coupling  consisting of $N=7$ nodes and 12 links. In both networks, each circle represents a logistic map. (c) TPRs and FPRs are plotted against various sample sizes $(T)$ for the network in Figure \ref{fig:directed_7_nodes}. (d) We also illustrated TPRs and FPRs 
    with respect to $(T)$ for the network in Figure \ref{fig:bidirected_7_nodes}. In both simulations, the number of permutations is $N_p =100$ and the significance threshold is $\theta=0.01$. }
    \label{fig:main}
\end{figure*}

\textit{Example 1: } In the first example, we choose the logistic map as a dynamical system in Equation \eqref{eq:network_equation} and its state equation is defined as
 
 \begin{equation}
  x_{n+1} = f(x_n) = a x_n (1-x_n) \quad x_0 \in [0,1].
  \label{eq:logistic_map}
 \end{equation}
Here, for $a = 4$ the system is chaotic. The number of nodes in the network is $N = 7$ in \eqref{eq:network_equation}, and we consider directed and bidirectional coupled networks as shown in Figures \ref{fig:directed_7_nodes} and \ref{fig:bidirected_7_nodes}. The coupling strength is $\sigma= 0.1$, and the inner coupling matrix is $\kappa= 1$. We choose the coupling function as $g(x_{n}^{(i)}, x_{n}^{(j)})= f(x_{n}^{(j)}) -  f(x_{n}^{(i)})$. The number of the permutation is selected as $N_p=100$ to determine the zero $\varepsilon_{forward}$, and the significance threshold is $\theta = 0.01$.  
The performance of the proposed algorithm is defined in terms of the True Positive Rate (TPR), the False Positive Rate (FPR) and the Receiver Operating Characteristic (ROC) curve \cite{sammut2011encyclopedia}. 

Figures \ref{fig:directed_tpr_fpr_7_nodes} and \ref{fig:bidirected_tpr_fpr_7_nodes} demonstrate the TPRs and FPRs with respect to various sample sizes (T) when the network is selected in Figures \ref{fig:directed_7_nodes} and \ref{fig:bidirected_7_nodes}, respectively.
As the number of samples increases, the TPR reaches one, and the FPR becomes zero as depicted in Figure \ref{fig:directed_tpr_fpr_7_nodes}. We could find all links correctly for the network in Figure \ref{fig:directed_7_nodes} using the $oGeoC$ algorithm. The TPR is almost one when $T$ is increased (i.e., the algorithm misses only one link as a false negative in Figure \ref{fig:bidirected_7_nodes}), and FPR drops when $T > 9000$,  as shown in Figure \ref{fig:bidirected_tpr_fpr_7_nodes}.

\textit{Example 2: } In this example, we investigate the case of randomly coupled networks. We use Erd\H{o}s-R\'enyi (ER) model to generate random graphs \cite{erdos1960evolution}, creating random couplings with a probability of p = 0.1 for the networks. The number of nodes is selected as  $N=20$. In particular, we choose the same dynamical system as the logistic map and the network parameters in Example 1. However, the number of independent trials is $10$. 

\begin{figure}[!ht]
    \centering
    \begin{subfigure}[b]{\textwidth}
        \centering
\includegraphics[scale=0.6]{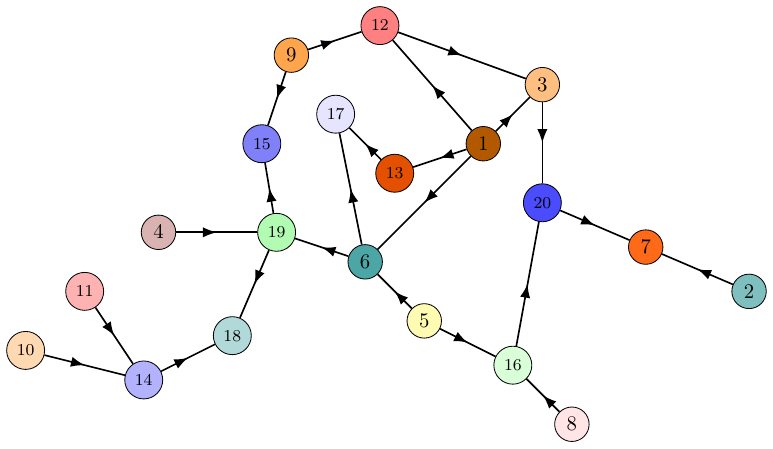}
        \caption{}
        \label{fig:twenty_nodes_logistic_network}
    \end{subfigure}
    
    \vspace{1em} 

    \begin{subfigure}[b]{0.45\textwidth}
        \centering  
   \includegraphics[scale=0.8]{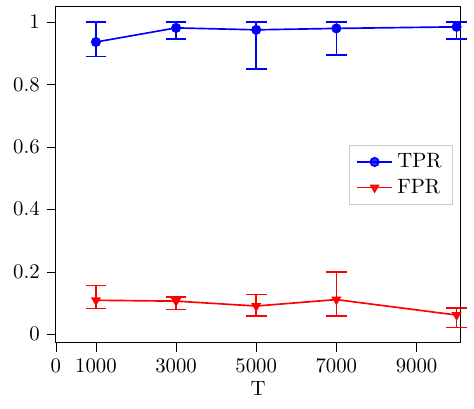}
        \caption{}
        \label{fig:tpr_fpr_over_10_trials}
    \end{subfigure}
    \hfill
    \begin{subfigure}[b]{0.45\textwidth}
        \centering 
        \includegraphics[scale=0.8]{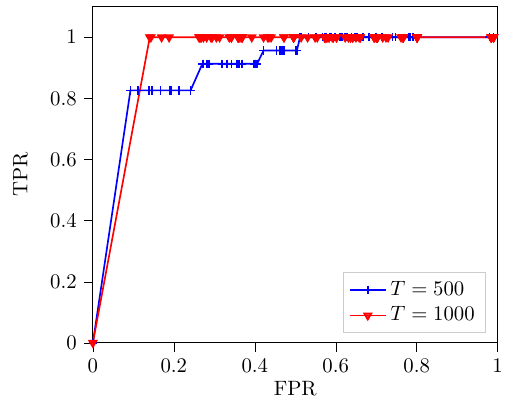}
        \caption{}
        \label{fig:roc_plot_N_500}
    \end{subfigure}
    
    \caption{The performance of the proposed algorithm  for the networks randomly coupled according to Erd\H{o}s-R\'enyi (ER) model with the probability of $p= 0.1$. The simulations are repeated 10 times for different networks (a) An illustration of one of the networks. (b) Error bar points show the mean of TPRs and FPRs with respect to different sample sizes. The maximum and minimum points of the error bar indicate the highest and lowest values of TPRs and FPRs. The number of permutations is again $N_p =100$ with $\theta= 0.01$. (c) ROC curve for $T=500$ and $T=1000$. Here, $N_p =100$, but  $\theta$ is varied from to $0.01$ to $0.99$ to plot ROC curve.}
    \label{fig:main_figure}
\end{figure}

We show only one of the realizations in Figure \ref{fig:twenty_nodes_logistic_network}. 
We illustrate TPRs and FPRs for different sample sizes in Figure \ref{fig:tpr_fpr_over_10_trials}. 
The points represent the average values of TPRs and FPRs over 10 trials. The maximum point of the error bar indicates the highest value of TPRs and FPRs, while the minimum point represents their lowest values. Additionally, the ROC curve is demonstrated in Figure \ref{fig:roc_plot_N_500}  with $T =500$ and $T=1000$ for the network in Figure \ref{fig:twenty_nodes_logistic_network}. The number of permutations is chosen as $N_p =100$. The significance threshold $\theta$ varies from $0.01$ to $0.99$ in Figure \ref{fig:roc_plot_N_500} to extract the ROC curve.

TPR approaches one, and the interval between the maximum and minimum values of the error bars is decreased when $T$ is increased, as shown in Figure \ref{fig:tpr_fpr_over_10_trials}. FPR is slightly reduced but not equal to zero for $T=10000$. In summary, the algorithm achieves to find true links in the networks; it detects a small number of links as false positives. 

The ROC curves in Figure \ref{fig:roc_plot_N_500} indicate that the performance of the proposed algorithms is improved when the number of samples is increased as expected. However, when $\theta$ is increased, the performance of algorithms decreases even if $N_p$ is large.

\section{Discussion}

In this study, we investigated the causal inference of networks using only geometric interpretations.  We utilized the conditional correlation dimensional geometric information flow measure based on the correlation dimension to accomplish this.
We proposed the $oGeoC$ principle, which allows us to find causal and noncausal parents of a node, thereby identifying direct and indirect links through geometric sense. We tested our proposed method on coupled logistic maps. Our findings revealed that we could find the links when the number of samples was large enough. False positives decreased when the observations were sufficient and the significance level $\theta$ was selected appropriately.

It is also important to note that the number of observations is a vital parameter when estimating $GeoC$. Although $GeoC$ detects causal relationships between systems, the estimation requires a large number of observations to estimate $\mathcal{D}_2$ accurately \cite{eckmann1992fundamental}.

We obtained $\mathcal{D}_2$ by finding the slope of $\ln C(\epsilon, T)/ \ln \epsilon $ curve over its linear regions. In this technique, the selection of the minimum radius $\epsilon_{min}$ and the maximum radius $\epsilon_{max}$ play key roles. If the dimensionality of the system is high, finding a linear region in the curve on high dimensional can be problematic even with a large number of observations. Hence, in case of large networks, it is important to increase the number of observations, and to select the radius $\epsilon_{min}$ and  $\epsilon_{max}$  
by considering linear region of  $\ln C(\epsilon, T)/ \ln \epsilon $ curve. 

Furthermore, we plotted ROC curves according to the significance level $\theta$ when the number of shuffles was fixed. If the number of shuffles is large enough, the $oGeoC$ algorithm can determine the causal links while removing the noncausal links. When the significance level $\theta$ is increased, the false positive rate reaches one, even if the number of shuffles is large, as the ROC curve depicts.

 As a direction for future work, it would be interesting to apply the $oGeoC$ principle to real data. In real data analysis, it may not be possible to observe all the states of the network nodes. In this case, it is necessary to reconstruct the phase space from a single observation of a node. The embedding dimension parameters (e.g., the delay of embedding and the embedding dimension) are significant factors when reconstructing the state space using Takens’ embedding theorem \cite{abarbanel2012analysis}. When the length of the data is sufficient, and there is no noise in the data, there exists a diffeomorphism between the reconstructed state space and the original space. It ensures the invariants of the system, such as $\mathcal{D}_2$, are preserved in the reconstructed phase space. However, the observations are generally too short or noisy in real data. Therefore, the choice of the embedding dimension parameters is essential for an accurate estimation of $\mathcal{D}_2$.

Determining $\mathcal{D}_2$ can also become challenging in noisy conditions.
The studies in \cite{kantz2003nonlinear, schreiber1993determination} have shown that we may not find the significant scaling region in the correlation sum and the linear interval of the correlation dimension
when the dataset has a 2\% noise level. Therefore, we may not obtain a reliable estimation of $GeoC$ in the presence of noise.

In addition, it is known that $\mathcal{D}_2$ estimation can be done with several techniques \cite{ji2011novel, krakovska2023simple, makarov2023correlation}. $oGeoC$ principle with different $\mathcal{D}_2$ estimation techniques or noisy data can be investigated for further exploration.

Although existing studies and our proposed method deal with  detecting causal inference in deterministic systems, the question of which method to use in stochastic systems, especially in weak and moderate coupling, remains unresolved. Another open question is whether different dynamic properties of interacting systems might bias the estimation of the causal direction. 
In a recent study, the ability of a state-space correspondence method to identify causal direction in nonlinear bivariate stochastic processes was investigated to solve these problems \cite{porta2024validity}. 

In our case, it is known that the $oGeoC$ principle utilizes the estimation of the correlation dimension. When the variance noise level of nonlinear bivariate stochastic processes increases, the estimation of the correlation dimensions for the systems becomes less reliable. As a result, it becomes difficult to identify causal and noncausal parents of a node and the performance of the proposed algorithms will be reduced accordingly. A potential solution to these challenges can be to examine the estimation of the correlation dimension using  newly proposed techniques in  \cite{ji2011novel, krakovska2023simple, makarov2023correlation} for nonlinear bivariate stochastic processes.

To conclude, we showed that the $oGeoC$ principle can detect the causal parents of a node in a network when the observations are long enough and entirely noise-free. It will be interesting to apply the $oGeoC$ principle to real data where the causal relations are unknown, including time series of air quality, temperature, and humidity. Future studies should also explore applications involving stochastic interactions,  such as determining causality in physiological control mechanisms, brain activity interactions, and coupled ocean-atmosphere chaotic systems. We aim to test this principle in a more detailed study and explore their causal relationships.  


\renewcommand{\thealgorithm}{A\arabic{algorithm}} 
\setcounter{algorithm}{0}

\renewcommand{\thefigure}{A\arabic{figure}} 
\setcounter{figure}{0} 
\appendix 
\section{Estimation of correlation dimension }
\label{sec:correlation_dim}

Given a time series $\left\{\mathbf{x}_{n} \right\}_{t=1}^T $, our goal is to estimate $\mathcal{D}_2$. Although $\mathcal{D}_2$ is invariant, the correlation sum
\eqref{eq:correlation_sum} is not invariant for a given $\epsilon$ \cite{kantz2003nonlinear}. Therefore, the correlation sum in \eqref{eq:correlation_sum} is calculated for several radii at first. Then, the curve of $\ln \hat{C}(\epsilon) / \ln(\epsilon)$ is plotted. The slope of the curve is computed over a linear region.  In this paper, we use the \textit{Linear Regression} model to determine the slope. The pseudo-code is given in Algorithm \ref{alg:corr_dim}. 

In our simulations, we choose the minimum radius as $\epsilon_{max}=0.0562$, the maximum radius as  $\epsilon_{max}=0.630$, and the number of radius steps as $\#_{rs} =50$ for the coupled logistic networks. We utilize \textit{kDTree} to count the pairs inside the ball of each $\epsilon$ and use \textit{multiprocessing} tools when estimating $\mathcal{D}_2$.

\begin{algorithm}[!ht]
\caption{$\mathcal{D}_2$ estimation}
\begin{algorithmic}[1]
\Procedure{$\mathcal{D}_2$ estimation}{$\mathbf{x}$, $\epsilon_{min}$,  $\epsilon_{max}$, $\#_{rs}$} 
\State \textit{Initialize:}  step\_$\epsilon$ = ($\epsilon_{max}$ -  $\epsilon_{min}$ ) /   $\#_{rs}$, 
\State $slope\_array \gets [\,]$
\State $radius\_array \gets [\,]$
 \For {$r=\epsilon_{min} ~\textbf{to} ~\epsilon_{max} $ \textbf{step} step\_$\epsilon$} 
\State Calculate  $\hat{C}(r)$ using $\mathbf{x}$
\State $slope\_array  \gets slope\_array  + [\ln \hat{C}(r)]$
\State $radius\_array  \gets radius\_array  + [\ln (r)]$
\EndFor
\State $(\mathcal{D}_2, residuals) \gets \text{Linear\_Regression}(radius\_array ,  slope\_array )$
\State \textbf{return} $\mathcal{D}_2$
\EndProcedure
\end{algorithmic}
\label{alg:corr_dim}
\end{algorithm}

\section{Shuffle test to determine the zero}
\label{sec:shuffle_test_to_determine_the_zero}
There are two thresholds, such as $\varepsilon_{forward}$ and $\varepsilon_{backward}$, to determine the zero in the $GeoC$ algorithm. We perform a statistical significance test to check whether $GeoC$ is greater than zero. The idea is to obtain an empirical cumulative distribution from the shuffled $GeoC$ and use it to determine the significance level. 

To achieve this, we generate a random permutation array that shuffles the time and surrogate $\mathbf{x}_n^{(j)}$ according to the permutation as  $\mathbf{x}_n^{(j^{\star})}$. Then, $GeoC_{j^{\star}~ \rightarrow ~ i ~|~ \mathcal{K} } $ is calculated. The procedure is repeated $N_p$ times, and the shuffled $GeoC$s are sorted in ascending order. The significance threshold $\theta$ 
and $N_p$ are used to set the predefined index, which determines the significance level. $\varepsilon$ is $predefined~index^{th}$ element of the ascending list of shuffled $GeoC$ values. The pseudo-code is given in Algorithm \ref{alg:shuffle_test}.

The determination of $\varepsilon$ depends significantly on the number of shuffles, $N_p$ and $\theta$. As the $N_p$ increases and $\theta$ decreases, we can determine zero significantly.

\begin{algorithm}[!h]
\caption{Shuffle test}
\begin{algorithmic}[1]
\Procedure{Shuffle Test}{$\mathbf{x}_n^{(i)}$, $\mathbf{x}_n^{(j)}$ $\mathbf{x}_n^{(\mathcal{K})}$ for  $n=1, 2, \ldots, T$ ; , $N_p:$ number of shuffles, $\theta$: significance threshold} 
\State \textit{Initialize:}  $array\_index \gets int(N_p \times (1 -\theta))$, 
\State $shuffle\_array \gets [\,]$
 \For {$trial \in N_p$} 
\State Generate a random permutation array  that shuffles n 
\State Shuffle $\mathbf{x}_n^{(j)}$ according the permutation and obtain shuffled times series $\mathbf{x}_n^{(j^{\star})}$
\State Calculate $GeoC_{j^{\star}~ \rightarrow ~ i ~|~ \mathcal{K} } $
\State $shuffle\_array \gets shuffle\_array + [GeoC_{j^{\star} ~ \rightarrow ~ i ~|~ \mathcal{K} } ]$
 \EndFor
\State $ascending\_Geo\_array \gets \text{sort}(shuffle\_array )$
\State $\varepsilon \gets ascending\_Geo\_array[array\_index]$

\State \textbf{return} $\varepsilon$
\EndProcedure
\end{algorithmic}
\label{alg:shuffle_test}
\end{algorithm}

\section{Illustration of $\mathcal{D}_2$ estimations  for the networks}
\label{sec:illustration_of_CD}
To demonstrate $\mathcal{D}_2$ estimations, we show  one step of the proposed forward algorithm for the networks in Figures \ref{fig:directed_7_nodes} and \ref{fig:twenty_nodes_logistic_network}, respectively.

First, we start with the network in Figure \ref{fig:directed_7_nodes}. Let us randomly choose the third node $(i=3)$ from the network. The forward algorithm calculates $GeoC_{j \rightarrow 3 ~|~ \emptyset}$ for all $j$ and returns the maximum value of $GeoC_{j \rightarrow 3 ~|~ \emptyset}$ for all $j$ and its index of this maximum value in the first iteration. In our case, we found that the maximum $GeoC$
as $max_{geoc}= GeoC_{2 \rightarrow 3 ~| ~\emptyset} = 0.163$ for the third node. When Equation \eqref{eq:geoc_emptyset} is written for $i=3$ and $j=2$, it becomes

  \begin{figure}[!ht]
    \centering
\includegraphics[scale=0.70]{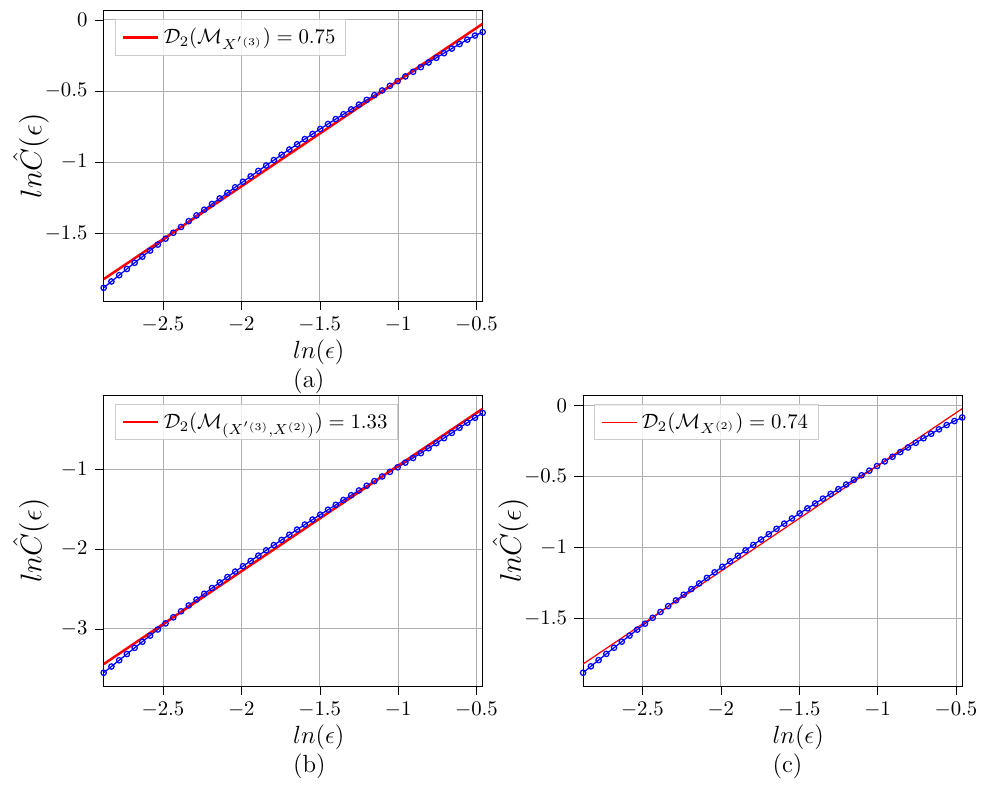}
\caption{The curve of $\ln \hat{C}(\epsilon) / \ln(\epsilon)$ for a given dataset of (a) $ X^{'(3)} $, (b) $(X^{'(3)}, X^{(2)})$, (c) $X^{(2)}$ for the network in Figure \ref{fig:directed_7_nodes}. The estimated correlation dimension is shown in the legend. 
The number of observations is $T=10000$.}
\label{fig:geoc12empty_23_empty}
\end{figure}

  \begin{figure}[!ht]
    \centering
\includegraphics[scale=0.70]{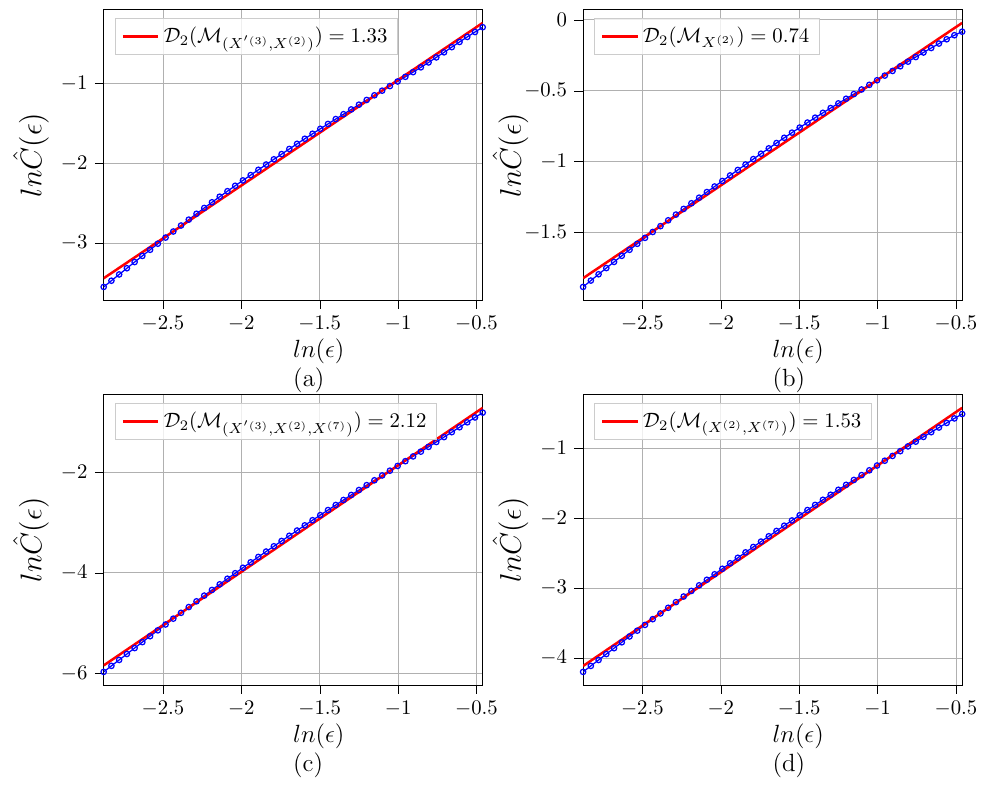}
\caption{The curve of 
$\ln \hat{C}(\epsilon) / \ln(\epsilon)$ for a given dataset of (a) $ (X^{'(3)}, X^{(2)}) $, (b) $X^{(2)} $, (c) $(X^{'(3)}, X^{(2)}, X^{(7)})$, (d) $(X^{(2)}, X^{(7)})$ for the network in Figure \ref{fig:directed_7_nodes}. The estimated correlation dimension is shown in the legend. 
   The number of observations is $T=10000$.}
\label{fig:geoc621_723}
\end{figure}

    \begin{equation}
    \begin{split}
    GeoC_{2 \rightarrow 3 ~| ~\emptyset} & = Geo(X^{'(3)}) - Geo(X^{'(3)}| X^{(2)}) \\
  &=  \mathcal{D}_2(\mathcal{M}_{X^{'(3)}}) -  \mathcal{D}_2(\mathcal{M}_{(X^{'(3)}, X^{(2)})}) + \mathcal{D}_2(\mathcal{M}_{X^{(2)}}).
    \end{split}
    \end{equation}

In the estimation of $GeoC_{2 \rightarrow 3 ~| ~\emptyset}$, we require to estimate the correlation dimension of $\mathcal{D}_2(\mathcal{M}_{X^{'(3)}})$, 
$\mathcal{D}_2(\mathcal{M}_{(X^{'(3)}, X^{(2)})})$ and $\mathcal{D}_2(\mathcal{M}_{X^{(2)}})$ for given datasets $ X^{'(3)} $,  $(X^{'(3)}, X^{(2)})$ and $X^{(2)}$, respectively. We use Algorithm \ref{alg:corr_dim} to estimate $\mathcal{D}_2(\cdot)$. We plot the curves of $\ln \hat{C}(\epsilon) / \ln(\epsilon)$ of the datasets  $X^{'(3)} $,  $(X^{'(3)}, X^{(2)})$ and $X^{(2)}$ in Figure \ref{fig:geoc12empty_23_empty}. The blue circle points demonstrate the natural logarithm of correlation sum versus the natural logarithm of radius.  The red line determines the slope of the blue points by using least squares estimation. In our case, the slope of the red line gives us the estimated correlation dimension.

In the algorithm, we determine whether the maximum value of $GeoC_{\cdot \rightarrow \cdot ~|~\cdot}$ is statistically significant or not. We found that $GeoC_{2 \rightarrow 3 ~| ~\emptyset} > 0  $. Hence, the index of the maximum $ GeoC_{j \rightarrow 3 ~|~ \emptyset}$ becomes $\mathcal{K} = index\_max =2$.

In the second iteration, we need to compute $ GeoC_{j \rightarrow 3 ~|~2}$ for all $j$ except for $\mathcal{K}=2$. Our results reveal that the maximum value of $GeoC_{j \rightarrow 3 ~|~2}$
    becomes $GeoC_{7 \rightarrow 3 ~|~2} = 0.001$ when $j= 7$. If $j=7$, $i=3$, and $\mathcal{K}=2$, Equation \eqref{eq:geo_j_i_k} is expressed by 

       \begin{equation}
    \begin{split}
    GeoC_{7 \rightarrow 3 ~| ~2} & = Geo(X^{'(3)} | X^{(2)}) - Geo(X^{'(3)}| X^{(2)},  X^{(7)}) \\
  &= \mathcal{D}_2(\mathcal{M}_{(X^{'(3)}, X^{(2)})}) 
  - \mathcal{D}_2(\mathcal{M}_{X^{(2)}}) -  \mathcal{D}_2(\mathcal{M}_{(X^{'(3)}, X^{(2)}, X^{(7)})}) + \mathcal{D}_2(\mathcal{M}_{(X^{(2)}, X^{(7)})}).
    \end{split}
    \end{equation}

Again, we illustrate the curves of $\ln \hat{C}(\epsilon) / \ln(\epsilon)$ and corresponding estimated correlation dimensions ( $\mathcal{D}_2(\mathcal{M}_{(X^{'(3)}, X^{(2)})})$,
$\mathcal{D}_2(\mathcal{M}_{X^{(2)}})$, $\mathcal{D}_2(\mathcal{M}_{(X^{'(3)}, X^{(2)}, X^{(7)})})$, and $\mathcal{D}_2(\mathcal{M}_{(X^{(2)}, X^{(7)})})$) in Figure \ref{fig:geoc621_723}.

In the second iteration, the maximum $GeoC$ is not statistically significant $(GeoC_{7 \rightarrow 3 ~|~2} < 0)$. Thus, the algorithm stops in the second iteration. We determine the causal parents of the third node as $\mathcal{K}=2$ for the network in Figure \ref{fig:directed_7_nodes} as expected.

\begin{figure}[!ht]
    \centering
\includegraphics[scale=0.7]{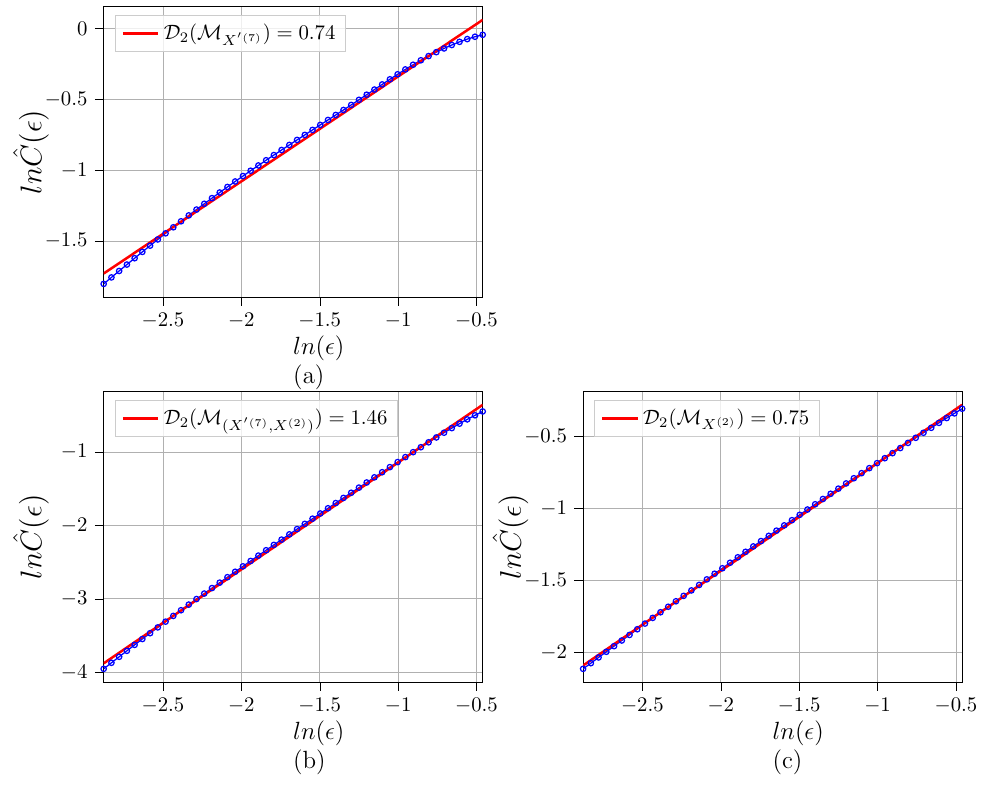}
\caption{The curve of $\ln \hat{C}(\epsilon) / \ln(\epsilon)$ for a given dataset of (a) $ X^{'(7)} $, (b) $(X^{'(7)}, X^{(2)})$, (c) $X^{(2)}$ for the network in Figure \ref{fig:twenty_nodes_logistic_network}.
The estimated correlation dimension is shown in the legend. The number of observations is $T=10000$.}
\label{fig:geoc16empty_20_node}
\end{figure}
\begin{figure}[!ht]
    \centering
\includegraphics[scale=0.7]{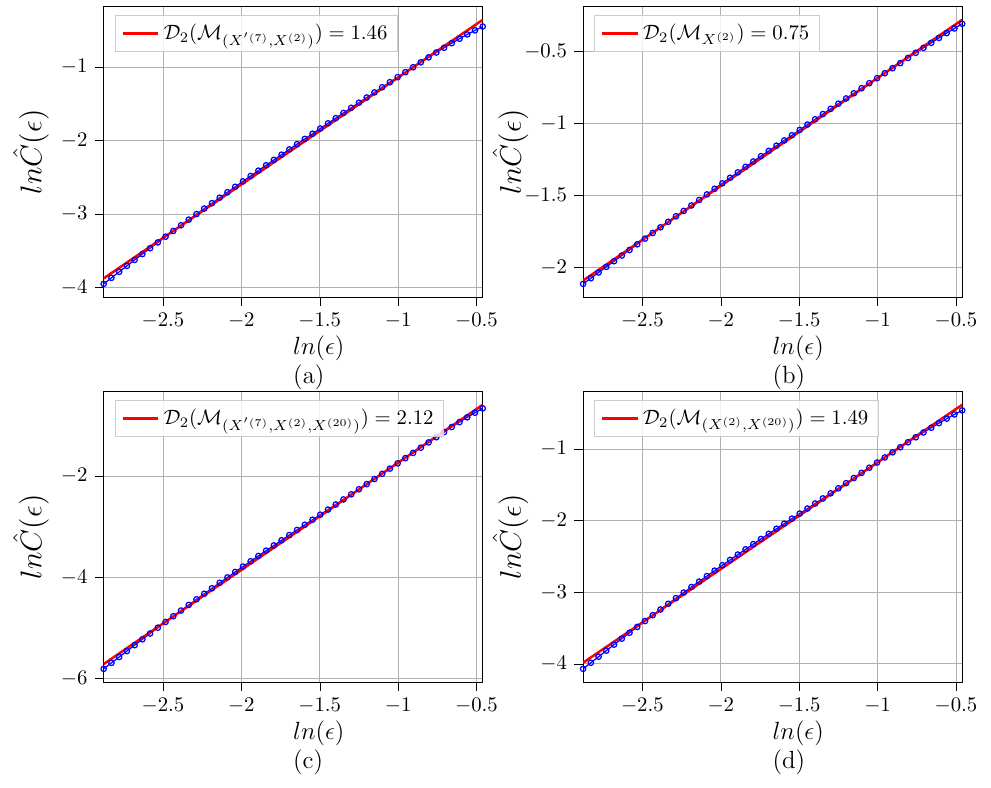}
\caption{The curve of 
$\ln \hat{C}(\epsilon) / \ln(\epsilon)$ for a given dataset of (a) $ (X^{'(7)}, X^{(2)})$, (b) $X^{(2)}$, (c) $(X^{'(7)}, X^{(2)}, X^{(20)})$, (d) $(X^{(2)}, X^{(20)})$ for the network in Figure \ref{fig:twenty_nodes_logistic_network}. The estimated correlation dimension is shown in the legend. 
   The number of observations is $T=10000$.}
\label{fig:geoc_1_6_19_20_node}
\end{figure}

Second, we continue with the network in Figure \ref{fig:twenty_nodes_logistic_network}. We randomly choose the seventh node $(i=7)$ from the network. 
The algorithm finds the maximum $GeoC$ as $GeoC_{2 \rightarrow 7 ~|~\emptyset}$ in the first iteration, and it is statistically significant ($GeoC_{2 \rightarrow 7 ~|~\emptyset} >0$). When estimating $GeoC_{2 \rightarrow 7 ~|~\emptyset}$, the algorithm computes the correlation dimension of the datasets $ X^{'(7)}$, $(X^{'(7)}, X^{(2)})$, and $X^{(2)}$ using the same technique  in Algorithm \ref{alg:corr_dim}. The curve of estimation of $\ln \hat{C}(\epsilon) / \ln(\epsilon)$ and the corresponding estimated $\mathcal{D}_2(\cdot)$ is shown in Figure \ref{fig:geoc16empty_20_node}.

In the second iteration, the maximum $GeoC$ is achieved
$GeoC_{20 \rightarrow 7 ~|~2}$. We demonstrate the curve of $\ln \hat{C}(\epsilon) / \ln(\epsilon)$ and the estimations of $\mathcal{D}_2(\cdot)$ of the dataset $ (X^{'(7)}, X^{(2)})$,  $X^{(2)}$, $(X^{'(7)}, X^{(2)}, X^{(20)})$, and $(X^{(2)}, X^{(20)})$ in Figure
\ref{fig:geoc_1_6_19_20_node}. The algorithm determines
$GeoC_{20 \rightarrow 7 ~|~2} >0$ and updates $\mathcal{K} = (2, 20)$ as a candidate set of causal parents. 

\begin{figure}[!ht]
    \centering
\includegraphics[scale=0.70]{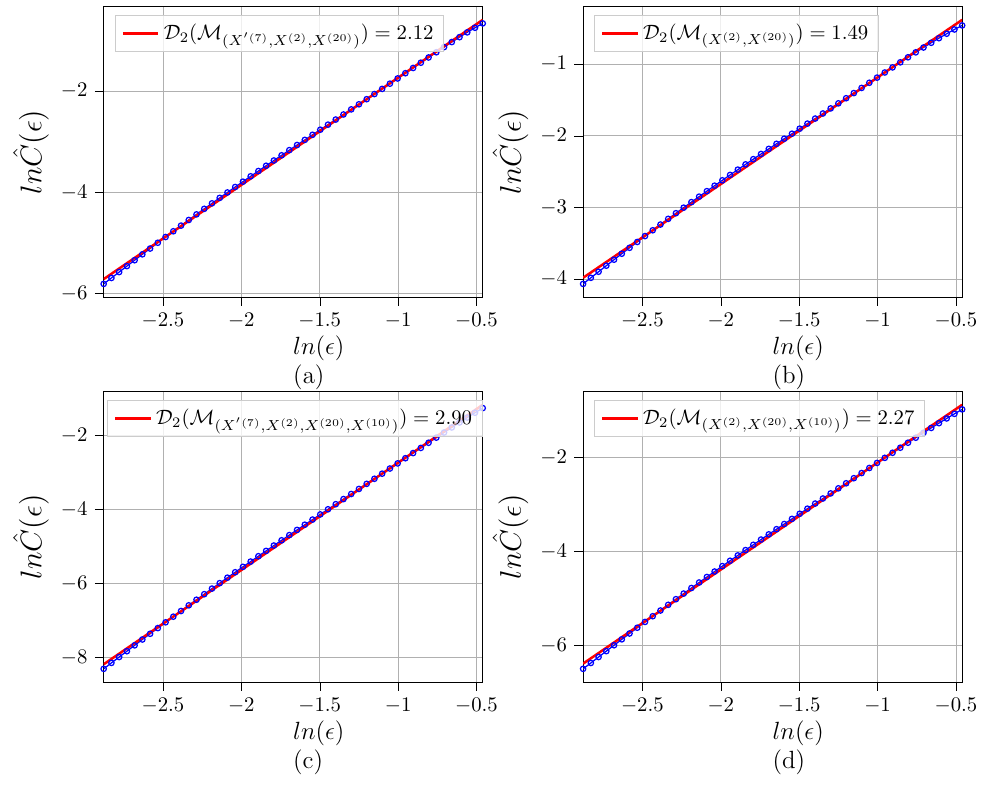}
\caption{The curve of 
$\ln \hat{C}(\epsilon) / \ln(\epsilon)$ for a given dataset of (a) $ (X^{'(7)}, X^{(2)}, X^{(20)})$, (b) ($X^{(2)}, X^{(20)}) $, (c) $(X^{'(7)}, X^{(2)}, X^{(7)}, X^{(10)})$, (d) $(X^{(2)}, X^{(7)}, X^{(10)})$ for the network in Figure \ref{fig:twenty_nodes_logistic_network}. The estimated correlation dimension is shown in the legend. The number of observations is $T=10000$.}
\label{fig:geoc1_6_19_9_20_node}
\end{figure}

In the third iteration, $GeoC_{10 \rightarrow 7 ~|~(2, 20)}$ takes the maximum value by calculating the correlation dimension of the datasets $(X^{'(7)}, X^{(2)}, X^{(20)})$, ($X^{(2)}, X^{(20)})$, $(X^{'(7)}, X^{(2)}, X^{(7)}, X^{(10)})$, and $(X^{(2)}, X^{(7)}, X^{(10)})$. The curves of $\ln \hat{C}(\epsilon) / \ln(\epsilon)$ of these datasets and their estimated $\mathcal{D}_2(\cdot)$
 are represented in Figure \ref{fig:geoc1_6_19_9_20_node}. In this case, the algorithm decides $GeoC_{10 \rightarrow 7 ~|~(2, 20)} < 0$ and returns the causal parents of the seventh node as $\mathcal{K} = \{2, 20 \}$.

\bibliographystyle{ieeetr}
\bibliography{sample}

\end{document}